\documentclass[12pt]{article}

\usepackage{scicite}

\usepackage{times}
\usepackage{graphicx}

\newenvironment{sciabstract}{%
\begin{quote} \bf}
{\end{quote}}

\newcounter{lastnote}

\date{\bf }

\title{Improved Spin-Dependent WIMP Limits from a 
Bubble Chamber}

\author
{E. Behnke$^{1}$, J.I. Collar$^{2\ast}$, P.S. Cooper$^{3}$, 
K. Crum$^{2}$, M. Crisler$^{3}$, M. Hu$^{3}$,\\
I. Levine$^{1}$,
D. Nakazawa$^{2}$, H. Nguyen$^{3}$, B. Odom$^{2}$, 
E. Ramberg$^{3}$,\\ J. Rasmussen$^{2}$,  N. Riley$^{2}$,
A. Sonnenschein$^{3}$, M. Szydagis$^{2}$ and R. Tschirhart$^{3}$\\
\\
\normalsize{$^{1}$Department of Physics and Astronomy,}\\
\normalsize{Indiana University South Bend, South Bend,
IN 46634, USA}\\
\normalsize{$^{2}$Department of Physics, Enrico Fermi Institute and Kavli Institute for Cosmological 
Physics,}\\
\normalsize{University of Chicago, Chicago,
IL 60637, USA}\\
\normalsize{$^{3}$Fermi National Accelerator Laboratory,}\\
\normalsize{Batavia, IL 60510, USA}\\
\\
\normalsize{$^\ast$To whom correspondence should be addressed; 
E-mail: collar@uchicago.edu}
}

\begin{document}

\baselineskip24pt

\maketitle

\begin{sciabstract}
   Bubble Chambers provided the dominant  particle detection 
   technology in 
   accelerator experiments for several decades, eventually falling 
   into disuse with the 
   advent of other techniques. We report here on the first period of 
   operation of an ultra-clean, room-temperature
   bubble chamber containing 1.5 kg of  superheated CF$_{3}$I,
   a target maximally 
   sensitive to spin-dependent and -independent Weakly Interacting 
   Massive Particle (WIMP) couplings. An exposure in excess 
   of 250
   kg-days is obtained, with a live-time fraction reaching 80\%. 
   This illustrates
   the ability to employ bubble chambers in a new realm, 
   the search for dark 
   matter particles. Improved limits on the spin-dependent WIMP-proton
   scattering cross section are
   extracted from this first period. An extreme intrinsic 
   insensitivity to the 
   backgrounds commonly limiting these experiments (a rejection 
   factor for photon-induced electrons of $\sim\!10^{-10}$) has been measured 
   in operating conditions leading to the detection of low-energy 
   nuclear recoils such as those 
   expected from WIMPs. 
\end{sciabstract}

\section*{Introduction}

With approximately 85\% of the total matter of the universe 
in a form which still eludes direct detection, the need for
large-mass, background-insensitive detectors 
able to explore the very small couplings expected from 
Weakly Interacting Massive Particles (WIMPs) is urgent.
While a member of this family of hypothetical particles,
the Lightest Supersymmetric Partner (LSP, \frenchspacing{ a.k.a. neutralino}) 
stands out as one of the most likely 
candidates for the  Dark Matter in galaxies \cite{wimps,revexp}, 
its predicted interaction rate via low-energy elastic scattering off 
nuclei
is $<<$1 event/kg-day for any target. 
In some unfavorable, yet entirely plausible models this can become 
a dismayingly 
small $<$1 event/ton-year.
To further aggravate this situation, the couplings still allowed by 
present direct searches  are 
already so weak that the next generation of massive (O(100)kg) WIMP 
detectors
will necessarily suffer from a penury of statistics in any  
dark matter signal they may observe. For instance, unique signatures such as a $\sim$5\% 
annual modulation in WIMP interaction rate expected from the orbital 
motion of the Earth \cite{andje} will soon require extreme exposures of 
O(100)Ton-year to become evident. Information 
about WIMP properties (e.g., mass of the particle) will be  scarce. This clearly 
indicates the need for a variety of detection techniques and targets, 
and for an emphasis within each experiment not only on improving
sensitivity, but on developing schemes that allow the  
unequivocal identification of these particles \cite{bertone}.

Superheated liquids, in the form of bubble chambers, were extensively 
used as the detection medium of choice
 in accelerator experiments spanning more than three decades. 
These chambers led not just to numerous discoveries, but also to an effect 
nearly as important: an unprecedented
ability to visualize fundamental particles and their interactions 
\cite{glaser,carlo}. The advent of newer technologies led to 
their progressive obsolescence during the seventies. 
In a bubble chamber, 
the heat deposited by ionizing radiations along 
their path produces local nucleations of the vapor
phase in a
delicate (metastable) superheated liquid. Rapidly growing
bubbles form along this path, 
are photographed, and the chamber is then reset by fast recompression 
to the stable liquid phase. Decompression follows, bringing the target below its vapor 
pressure at the operating temperature, sensitizing it to radiation 
and starting the cycle anew. 
In accelerator experiments it was sufficient to maintain the 
superheated state during the few milliseconds corresponding to the 
bunched arrival of incident particles. Uncontrollable boiling on porous materials 
(gaskets, metallic surfaces) prevented serious consideration 
of this type of detector in searches for rare events, where the time of arrival of the 
signal is not known. 

Taking several precautions towards the deactivation of inhomogeneous 
bubble nucleation 
centers, we have recently 
shown that it is possible to achieve what is in principle an
indefinitely long stability in moderately superheated 
bubble chambers \cite{prl2}, 
thereby allowing their 
use in direct WIMP searches. In this realm of application, several 
 advantages 
for this type of detector can be mentioned. Most 
important is the fact that 
the superheated liquid can be tuned 
to respond exclusively to particles having a large 
 stopping power ($dE/dx$, energy loss per unit path distance). 
 In this way muons, gamma rays, x-rays, beta particles, etc., all fall 
well below a bubble nucleation threshold, which is typically   
$dE/dx>50~keV/\mu m$ during a WIMP 
search ({\frenchspacing Fig. 1}). 
The target liquids are  nevertheless 
sensitive  to few-keV nuclear recoils such as those expected from 
the scattering of WIMPs, given their much denser energy 
deposition. The process of radiation-induced bubble nucleation is described within the 
framework of the classical ``Hot Spike'' model \cite{seitz}: for the phase transition to occur, the  energy  deposited
by the particle must be larger than the thermodynamical
work of formation of a critically-sized protobubble, but this 
energy must also be lost over a distance comparable to the size of 
this protobubble, 
i.e., a minimum stopping 
power condition must be fulfilled. This condition leads to an advantageous insensitivity to 
the listed minimum ionizing backgrounds that normally 
plague WIMP searches ({\frenchspacing Figs. 1,2}). 
Formal details on 
the theory of radiation-induced bubble formation in superheated 
fluids can be found in 
\cite{myprd,prlsdd1,prlsdd2} and references therein. 
As for the probability of spontaneous (homogeneous) bubble nucleation in 
the bulk of a superheated fluid, it only becomes sizeable a few degrees below 
the critical temperature of the target compound \cite{homog}. In 
normal operating conditions for a dark matter search this source of 
instability is entirely 
negligible.

\section*{Experimental}

\paragraph*{Chamber operation}

Early in 2005 we installed a chamber containing 1.5 kg of  
CF$_{3}$I, an industrial refrigerant commonly used as a fire 
extinguisher, at the 350-foot
depth of the NuMI (Neutrinos at the Main Injector) tunnel in 
Fermilab. 
Environmental 
neutrons 
can produce single bubbles such as those expected from WIMPs if they scatter just once
in the target ({\frenchspacing Fig. 2}). At this shallow depth, a 
combination of active and passive shielding can  
lead to a neutron background rate 
as low as 0.01 
bubbles / kg-day. The choice of  
CF$_{3}$I provides 
optimal sensitivity to both spin-dependent (SD) and 
spin-independent (SI) WIMP couplings, 
in the first case by its 
fluorine content, in the second via the presence 
of iodine \cite{wimps,revexp,SDSI}. 
This allows a maximally efficient 
exploration of supersymmetric WIMP candidates \cite{bertone}.

Bubble production within the chamber is 
monitored via two trigger mechanisms, the sound emission or 
pressure rise created by their formation and 
growth, and by the changes that their appearance induces 
in CCD camera 
images, inspected every 
$\sim$20 ms. Two orthogonal cameras watch the inner 
 volume ({\frenchspacing Fig. 2}), where the 
 active liquid resides within a thin quartz 
vessel. A bellows mechanism balances the pressure difference across 
the quartz vessel wall \cite{hilde}. The precision in the stereoscopic 
reconstruction of the spatial position of 
small early bubbles is of the order of their size at the time of the 
trigger, $\sim$1 mm. A live-time fraction of $\sim$80 \% was obtained after the adoption of 
real-time image analysis as the primary trigger. 

The chamber was
operated continuously, in a variety 
of conditions (pressure, temperature, presence of radioactive 
sources, different triggering methods and pressure cycling protocols), 
for a year starting on 
December 2005, generating an exposure larger than 250 
kg-days, with an overall mean time between expansions of $\sim$100 s. 
The emphasis on this first large prototype was purely on the functional aspects 
leading to continuous unattended operation and enhanced stability, obtained 
by implementing the 
techniques described in \cite{prl2}. In the interest of 
rapid deployment, little attention was 
paid to alpha-recoil backgrounds such as those induced by radon and 
its progeny
({\frenchspacing Fig. 1}). The data 
at hand, although entirely dominated by
radon-induced backgrounds, already yield
an improved sensitivity to SD WIMP-proton couplings.

\paragraph*{Calibrations}

Of importance prior to a WIMP search is an empirical determination of
the maximum degree of superheat \cite{dsh} achievable
before the 
onset of sensitivity to minimum ionizing particles such as 
photoelectrons ({\frenchspacing Fig. 1}). This in turn defines the lowest 
recoil energy that can be detected in background-free conditions, and 
with it the acceptance for a WIMP 
signal. Similarly, it was 
necessary to demonstrate that sensitivity to low-energy 
nuclear recoils was nevertheless present in those conditions. Two 
calibration sources were developed for these purposes. The first one, 
an intense 13 mCi 
$^{137}$Cs gamma source, was placed next to the chamber's outer steel 
vessel, inside 
its  neutron-moderating polyethylene
shield. {\frenchspacing Fig. 3} shows the response of the chamber as a function of 
operating pressure in the presence of the source and in its absence. A 
Monte Carlo simulation \cite{mcnp} was used to calculate the rate of 
photon-induced electron production within the active 
volume of the chamber.
From the difference between this interaction rate 
and the observed bubble nucleation rate in 
the presence of the source it is possible to
obtain a gamma rejection factor (the fraction of interacting gammas 
inducing bubbles) as a function of degree of superheat. 
This factor can then be expressed in terms of 
the calculated nuclear recoil energy threshold for 
each pressure setting ({\frenchspacing Fig. 
3}, inset). The minimum-ionizing 
background rejection obtained, $\sim\!10^{-10}$ for a nuclear 
recoil threshold of $\sim$10 keV, is unmatched by any other WIMP 
detector. For instance, dark matter detection efforts using 
cryogenic germanium detectors sensitive to both ionization and heat 
(Akerib {\it et al.} in \cite{SDSI}) presently feature a rejection factor
$10^{-4}\!-\!10^{-5}$ for a similar recoil threshold. As a 
result of this, moderately superheated bubble chambers are
free from any
radiopurity constraints in gamma or beta emitters in the detector or its 
neutron moderator shield, including possible elevated rates of $^{14}$C in 
the active liquids. This freedom, when combined with room-temperature 
operation, translates into large gains in construction speed and a
reduction in costs. This background insensitivity is {\it intrinsic}: 
electron-induced events simply do not take place.

The second calibration source is 
a switchable Am/Be neutron source ({\frenchspacing Fig. 4}). It was 
characterized using a $^{3}$He counter surrounded by 
moderator. A Monte Carlo simulation \cite{mcnp} was used to generate the 
response of the counter, leading to a measured neutron yield 
of 4.9 n/s, in excellent agreement with 
the $(\alpha,n)$ reaction yield from the $^{241}$Am foil sources 
employed and separately characterized. In order to further
assess the uncertainty in the very small yield of the source 
\cite{suff}, the same simulation package and neutron
detector were used to characterize five commercial neutron 
sources
of known activity. 
This uncertainty was 
found to be a modest $\pm$11\%.
This switchable source, also placed  outside of the 
steel wall of the recompression chamber, is 
moderated by the $\sim$10 cm of 
recompression fluid around the quartz vessel and is therefore
expected to produce recoils very similar in energy to those  
from WIMP interactions ({\frenchspacing Fig. 4, inset}). 
In comparing the predicted response 
to this source with actual observations, one must take into account 
additional sources of uncertainty affecting the 
input to the MCNP-PoliMi Monte Carlo \cite{polimi} used to 
transport neutrons and  
generate the expected rate and 
energy distribution of recoils in the active liquid. We have appraised 
these to be dominated by a $\pm$28\% from the fiducialization of the 
chamber geometry and a $\pm$24\% from the hardness of the neutron spectrum 
(affected by alpha-particle energy losses on their way to the Be 
foils \cite{alp}).
The comparison is very good, as 
can be appreciated in {\frenchspacing Fig. 4}. These predictions were 
generated prior to inspection of the neutron
irradiation data. 
The success of this 
calibration confirms the
 predictions for bubble nucleation thresholds as a function of 
 superheat, 
sanctioning WIMP 
limits obtained from the device. Switchable sources similar to this one can be used 
in planned larger chambers for periodic studies of response to 
nuclear recoils, 
important for instance when looking for WIMP-induced modulations in 
the data \cite{andje}. To the best of our knowledge this is the first instance of 
their use in a direct search for WIMPs. 

\paragraph*{Present limitations of the method}

Two backgrounds became evident during underground operation of the 
chamber. The first is an excess of bubbles 
on the wall of the quartz vessel. 
A fraction of their rate may arise from the exposure of the vessel to 
typical concentrations of radon in air before installation. This 
results in the
shallow implantation of its long-lived alpha-emitting daughters \cite{implant}. 
Another source able to explain most to all of the observed surface-alpha rate is the 
$\sim$50 ppb of uranium measured 
via gamma spectroscopy in the quartz vessel material.  An origin in alpha 
emission for these wall events is confirmed by their 
disappearance at pressures higher than 65 (35) psig, the predicted 
thresholds for alpha-induced bubble nucleation
when operating at 40 (30) C. Image 
analysis readily identifies these bubbles as 
happening on the wall, allowing rejection with high 
confidence. However, since each event is followed by tens of seconds 
of recompression time to ensure recovery of the superheated state in 
the subsequent decompression \cite{prl2}, an excessive wall rate per unit quartz 
surface 
would lead to a considerable dead time in larger chambers. 
A decrease by $\times$10 ($\times$100) in the wall event rate
from the presently observed value is needed 
to ensure that 
O(50) kg 
chambers under construction sustain a live time $>$60 \% ($>$85 \%). In order to achieve this 
reduction, an etching procedure has been developed in collaboration with the vessel
manufacturer, facilitating the removal of 
implanted daughters while preserving surface smoothness and 
resistance against fracture initiation.  In addition, new chambers 
employ synthetic fused silica containing only a few ppt of alpha-emitters.

The second source of background events from this mechanical prototype 
detector consists of 
temperature-dependent radon emanations and injections
into the chamber, likely from 
thoriated weld lines and a Viton O-ring exposed to the inner detector 
volume. Viton is known to release radon at a rate of
$\sim$300 mBq/$m^{2}$ and does not create a strong barrier against 
its diffusion from external sources. Radon decays can lead to single 
bubbles in the bulk
identical to those expected from WIMPs (they nevertheless display a very different 
spectral and temporal dependence, as discussed below). That the origin of 
essentially all 
presently-observed
bulk events is in 
radon-associated alphas and alpha-recoils is evidenced by two signatures. First, the distribution of 
times between consecutive bulk events ({\frenchspacing Fig. 5}) displays a 
clear short-lived component corresponding to the  
the decay 
sequence 
$^{222}\!Rn\rightarrow^{218}\!\!\!Po\rightarrow^{214}\!\!\!Pb$ (both 
steps are alpha-emitting). Two independent analyses of this 
distribution, 
including the full decay chain ({\frenchspacing Fig. 5}), the effect of recompression (dead) 
time, fiducial volume cuts and the possibility of other sources being present, strongly 
favor models with $\sim$100\% of the events belonging to the 
$^{222}$Rn
sequence and a $\sim$100\% bubble nucleation efficiency for 
the combined alpha and alpha-recoil emission. Second, 
the distribution of bulk event rates vs. operating pressure at a 
fixed temperature is flat below a sharply defined
onset pressure, the value of which is again in good agreement with 
theoretical expectations ({\frenchspacing Fig. 5}). 
This is the kind of spectral behavior expected from the response of a 
threshold detector to a 
mono-energetic source. It has 
been observed in the past in bubble chambers intentionally spiked 
with alpha emitters
\cite{poes}. New larger
chambers under construction use metallic seals, with all weld 
lines being non-thoriated, among other precautions against Rn. 

\section*{Improved sensitivity to WIMP couplings}

Even constrained by the background limitations of the mechanical prototype in these early 
runs, an improved 
sensitivity to the spin-dependent WIMP-proton coupling is obtained from 
these data. 
This is possible due to the extreme sensitivity to this mode of 
interaction afforded by the large mass fraction 
of fluorine (29.1\%) and the operation of the chamber in conditions where 
only nuclear recoils (alpha, neutron or WIMP-induced) can produce bubbles. 
{\frenchspacing Fig. 5} displays the characteristic recoil rate
expected from example WIMPs, overlaid on the data.
Such spectra are calculated for different WIMP masses using the 
theoretical bubble nucleation 
thresholds (top axis in the figures) for fluorine recoils \cite{iod}. These are in turn
used to generate the integrated WIMP recoil rate above 
threshold, following the 
method and dark matter halo parameters recommended in \cite{smith2}. 
These threshold calculations have been validated by their success in predicting the 
response of the chamber to the Am/Be source \cite{soft}. 
Following a procedure similar
to that in \cite{picasso}, experimental bulk bubble nucleation rates vs. 
pressure, such as those depicted in {\frenchspacing Fig. 5}, are fitted with a 
model containing two free parameters: 
the signal from a WIMP of a given mass, scalable by a free spin-dependent 
cross-section, and the 
response to $^{222}$Rn alphas, expressed as a logistic function with a
free overall normalization ({\frenchspacing Fig. 5}). 
The minimization is performed using the MINUIT package\cite{minuit}, 
its output cross-checked via a standard error matrix 
analysis. 
The 
best fits favor the null hypothesis alone (response to $^{222}$Rn
alphas only). The largest spin-dependent couplings allowed by the 
data, up to a 90 \% C.L., are displayed in the form of exclusion plots 
in {\frenchspacing Fig. 6}. These bounds arise from the weighted average \cite{smith2}
of the analyses performed on three independent large data sets adding up 
to 52 kg-days of exposure.

\section*{Conclusions}

Even prior to any measures against alpha-emitting backgrounds, improved 
bounds on spin-dependent WIMP couplings can be obtained from the 
application of an old technology, bubble chambers, to a new problem, 
that of dark matter 
detection. In particular, these new bounds make a spin-dependent 
interpretation \cite{DAMA1} of the DAMA claim of an observed WIMP 
signal \cite{DAMA2} very difficult. Other experiments have already 
severely constrained spin-independent explanations.

In principle, and strictly from the point of view of internal 
backgrounds, the sensitivity of this technique can be improved by 
as much as six 
orders of magnitude by reducing 
alpha-emitting contaminants from the uranium 
and thorium chains to levels 
similar to the best achieved in large neutrino experiments  
($<10^{-17}$ g/g) \cite{jap}. This would guarantee an extensive probing of 
supersymmetric WIMP models along both spin-dependent and -independent 
couplings \cite{bertone}. The possibility exists to add  
the detection of dark matter 
to the many contributions already made by Bubble Chambers to particle 
physics. 
We presently concentrate on the development of radon-free chambers 
with a total target mass of O(50) kg.

We gratefully acknowledge the effort and outstanding technical support of
the Fermilab Staff.  This work is supported by NSF CAREER award 
PHY-0239812, NSF grant PHY-0555472, the Indiana
University South Bend R\&D committee,
the Kavli Institute for Cosmological Physics through grant NSF 
PHY-0114422, and the U.S. Department
of Energy.

\begin{figure}
\centering
\includegraphics[width=14cm]{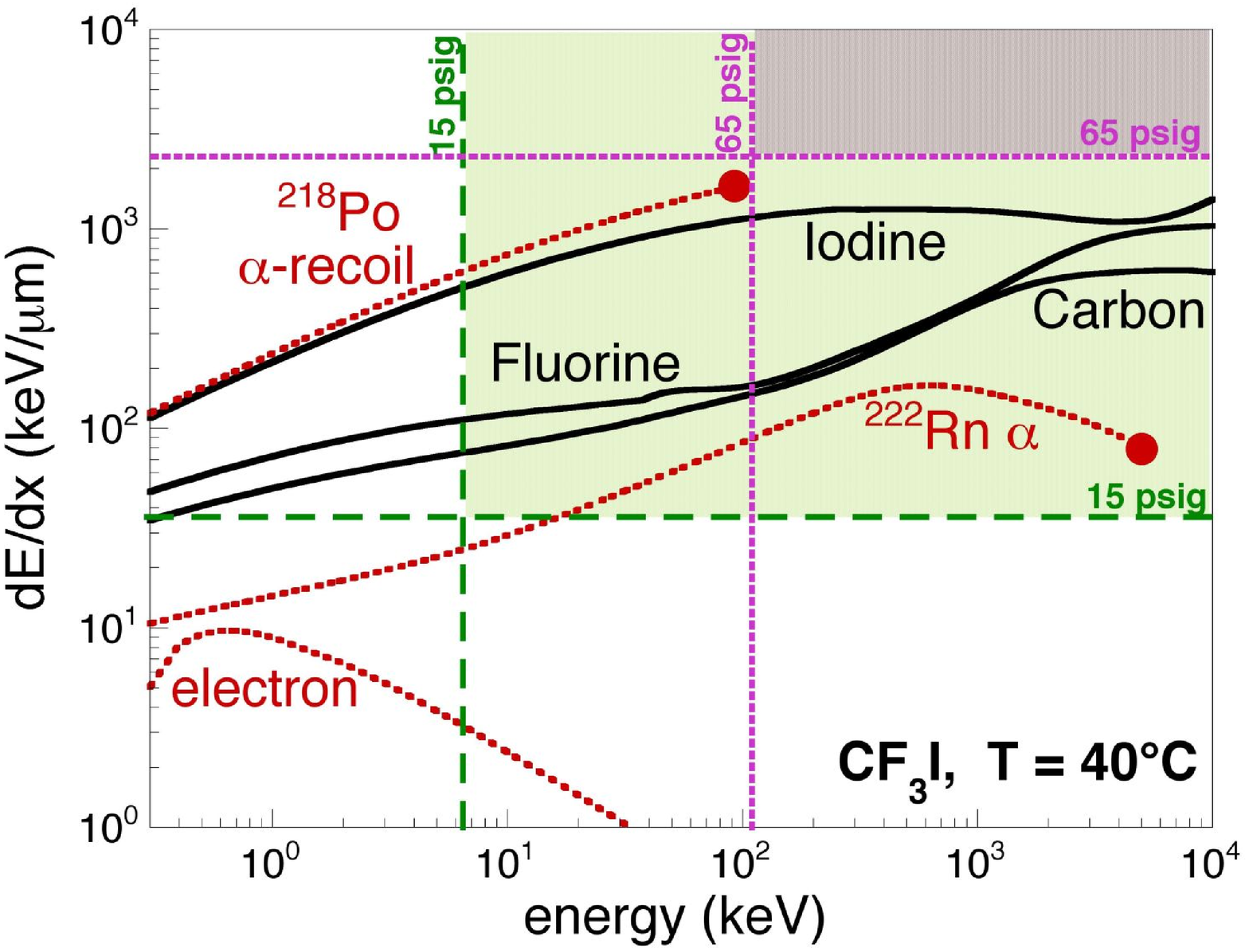}
\caption{\label{fig:epsart}Instantaneous stopping power vs. energy for 
different particles in CF$_{3}$I, including its three recoiling species. 
Plotted as horizontal and vertical lines are the calculated dE/dx and 
energy 
thresholds for bubble nucleation at T=40 C and two different 
operating pressures. According to the ``Hot Spike'' nucleation model
\protect \cite{seitz} only radiations in the 
top right (colored) quadrants can 
lead to bubbles. Notice the absence of this possibility for 
electrons even toward the end of their range, in conditions that 
nonetheless lead  to a  sensitivity to  recoils of just a few keV, 
such as those expected from WIMP 
interactions. Alpha particles and their recoiling daughters can 
induce bubble nucleations and the presence of their emitters must therefore 
be avoided. }
\end{figure}

\begin{figure}
\centering
\includegraphics[width=14cm]{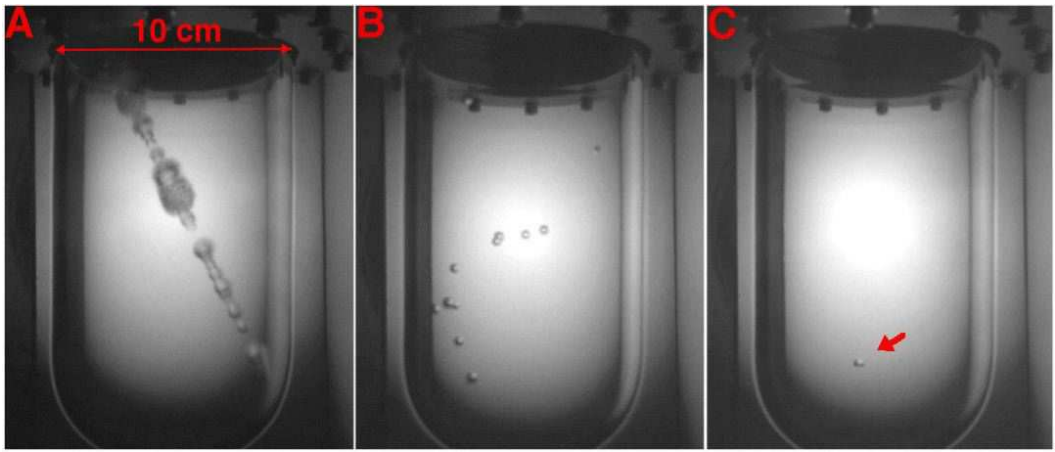}
\caption{\label{fig:epsart}Families of events in a 1.5 
kg CF$_{3}$I bubble chamber. At high degrees of superheat ($\sim$60 C 
and atmospheric pressure for this compound), minimum ionizing cosmic ray events 
reminiscent of those observed in early bubble chambers \protect\cite{glaser} are 
visible (left). At moderate superheats ($\sim$30 C, 1 atm) 
the chamber is sensitive strictly to high $dE/dx$ radiation such as 
nuclear recoils. Whereas neutrons can give rise to simultaneous 
separate bubbles each corresponding to a scatter (center), WIMPs are 
expected to produce single bubbles only (right), due to their extremely 
small probability of interaction. The mean free path between scatters 
is of just a few cm for neutrons, leading to an excellent ability to 
reject them in large chambers.
}
\end{figure}

\begin{figure}
\centering
\includegraphics[width=14cm]{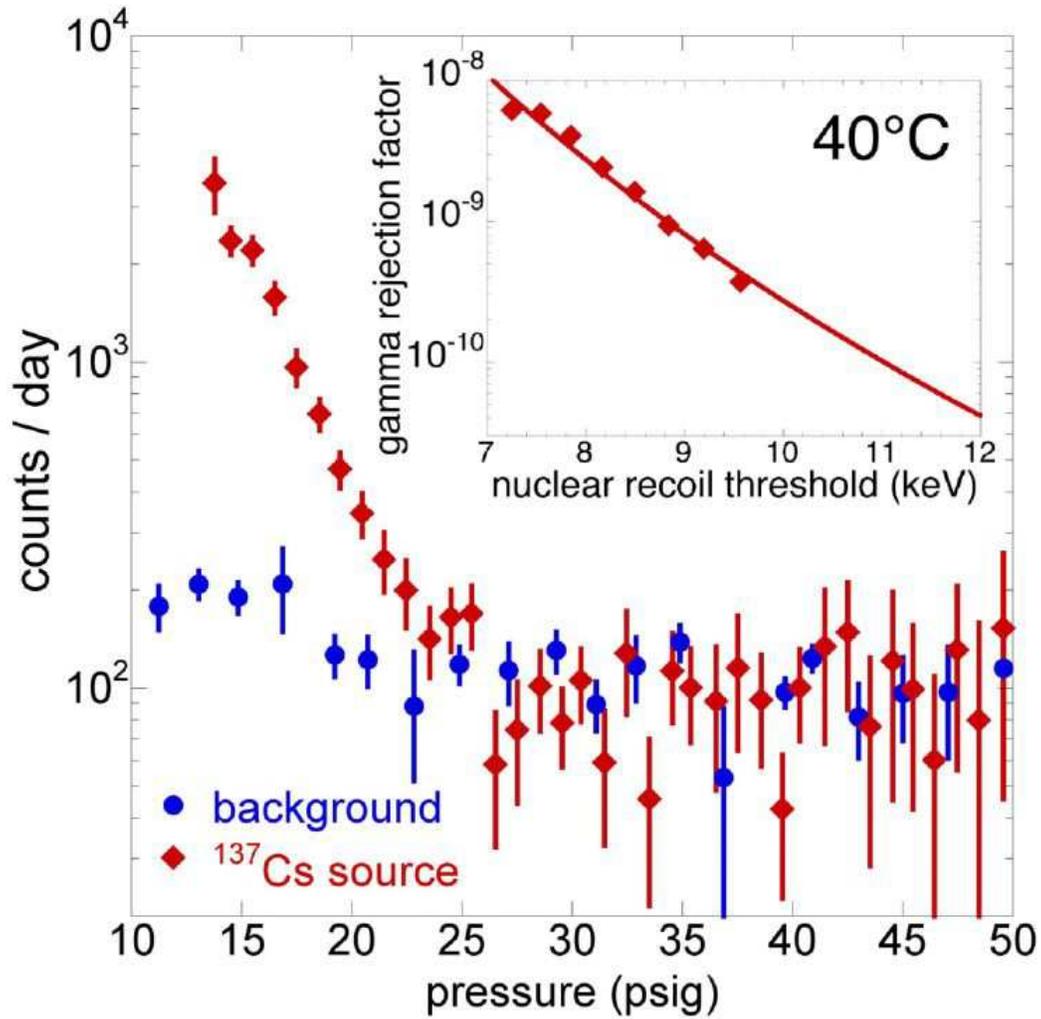}
\caption{\label{fig:epsart} Response 
of the chamber to an 
intense $^{137}$Cs gamma source. The expected
gamma interaction rate with the superheated liquid is 3.9$\times10^{6}$ per second, 
with gamma-induced electron energies reaching up to 662 keV.
{\it Inset}: Intrinsic gamma rejection factor (fraction of interacting gammas 
inducing bubbles)
obtained from the exposure to the source (see text).}
\end{figure}

\begin{figure}
\centering
\includegraphics[width=14cm]{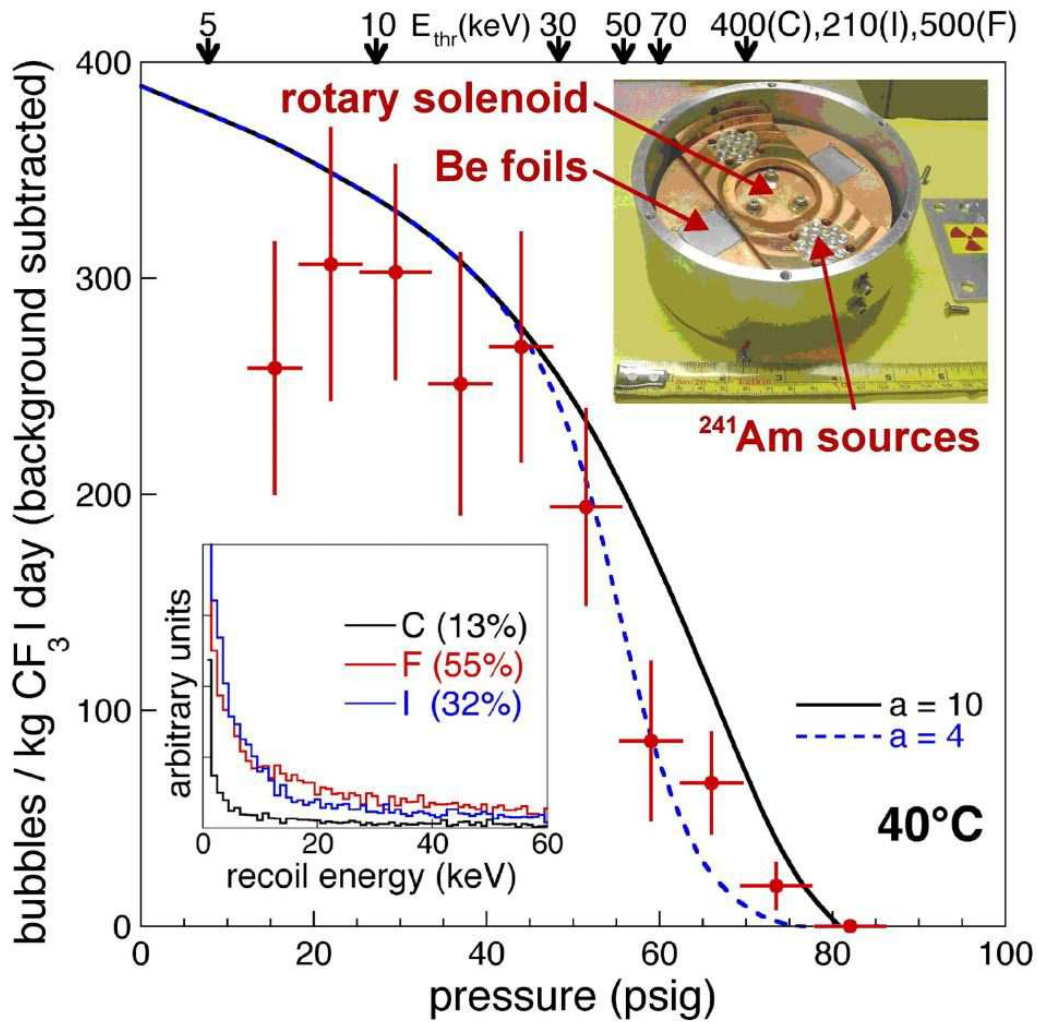}
\caption{\label{fig:epsart} 
Blind absolute comparison between expected bubble nucleation rate (lines) 
and observations (points) in the active presence of the switchable Am/Be 
neutron source ({\it top inset}). 
The two lines correspond to largely different values of a nucleation 
parameter ``a'' \protect \cite{jap2}, the single free factor in 
the classical theory used to predict bubble nucleation thresholds 
\protect \cite{seitz}. 
A fit to 
the data provides an excellent agreement with theoretical expectations 
(a = 6) \protect \cite{bell}. 
Including the uncertainty in the predictions (see text) the 
same fit simultaneously yields an efficiency in the response to this source of 
$81\pm^{63}_{33}$\% ($51\pm^{40}_{18}$\%) at 40 C (30 C) (errors are 
90\% confidence levels). Calculated thresholds for recoil-induced 
nucleation are expressed in keV along the top axis.
{\it Bottom inset:} Spectrum of nuclear recoil energies 
produced by the source (MCNP-PoliMi 
simulation \protect \cite{polimi}), similar to typical expected WIMP 
recoil spectra.}
\end{figure}

\begin{figure}
\centering
\includegraphics[width=12cm]{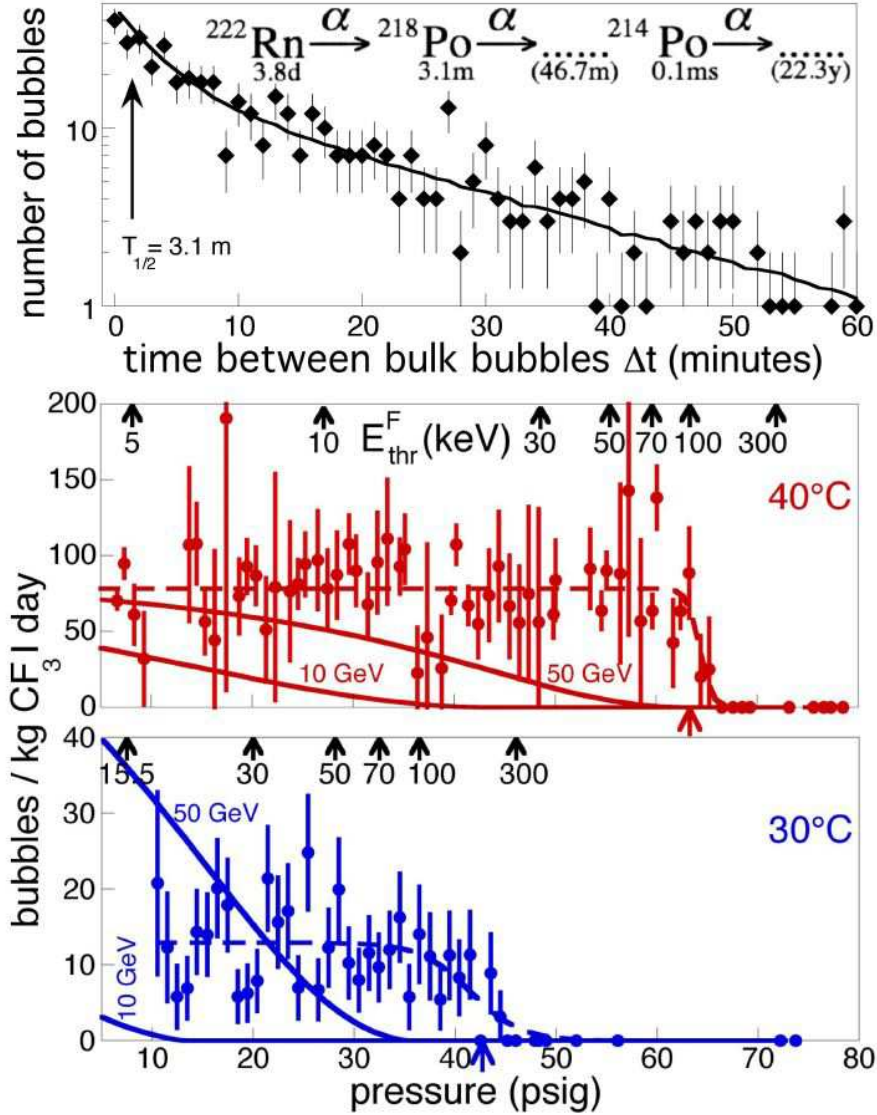}
\caption{\label{fig:epsart} 
{\it Top:} Distribution of times between 
consecutive bulk events in 
the chamber. A model (solid line) including the effect of 
dead time and fiducial volume
cuts, based on the triple alpha emission from 
$^{222}$Rn and progeny, successfully reproduces this distribution (see text). 
{\it Bottom:} Distribution of single bulk bubble 
nucleation rate vs. operating pressure, for two 
different running temperatures.
Rates display a flat behavior up to a pressure endpoint. This is 
characteristic 
of the response to monochromatic alphas and $\sim$100 keV alpha 
recoils (here from $^{222}$Rn 
emanations, see text). Colored arrows along the bottom 
axes indicate the predicted onset of 
sensitivity to these particles, in good agreement with 
observations. As a reference, solid lines correspond to the 
expected signal rate from WIMPs with a mass of 10 and 50 GeV/$c^{2}$ and a 
3 pb cross section for their spin-dependent coupling to protons. Also 
shown is the response function to $^{222}$Rn and 
progeny 
(dashed lines).
Calculated energy thresholds (in keV) for bubble nucleation by fluorine 
recoils are shown along 
the top axes. }
\end{figure}

\begin{figure}
\centering
\includegraphics[width=10.5cm]{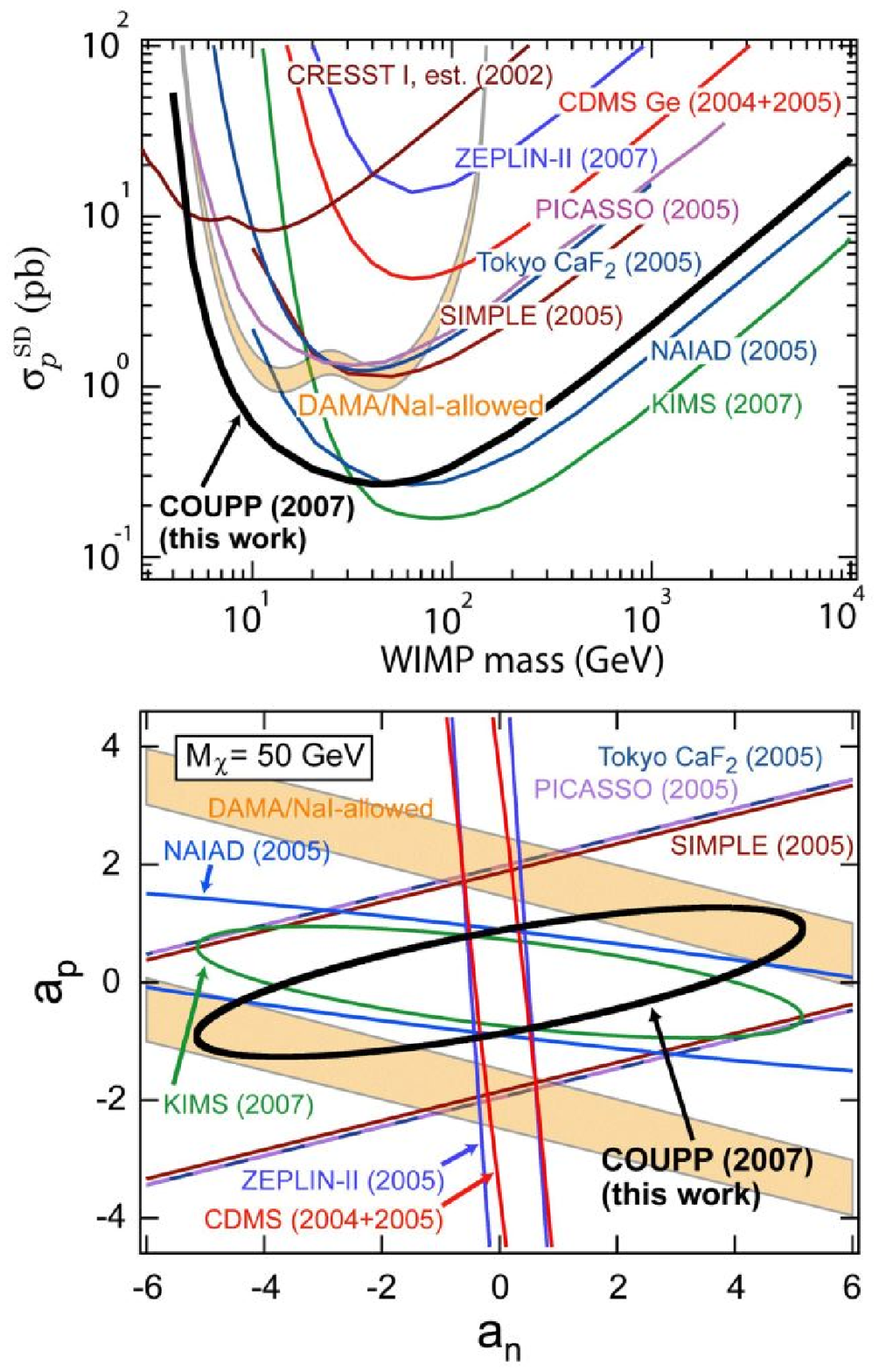}
\caption{\label{fig:epsart}  
{\it Top:} 
Improved limits on spin-dependent (pure) proton-WIMP coupling vs. WIMP 
mass from this experiment (COUPP, 
the Chicagoland Observatory for Underground Particle Physics). 
Couplings above the line
would have produced signals above observed backgrounds and are 
excluded to 90\% C.L. Limits from other experiments are also shown 
\protect\cite{ricky}, 
as well as the (orange) region favored as a possible explanation to 
an existing
claim for WIMP observation \cite{DAMA1,DAMA2}, a hypothesis now
contradicted by this experiment.
{\it Bottom:} Similar limits for spin-dependent coupling parameters 
where no assumption is made about the relative strength of the 
coupling to neutrons and protons, but a WIMP mass must be chosen (50 
GeV/c$^{2}$ here)
\protect\cite{franco1,franco2}. The 
region outside of the ellipses is excluded by each experiment.}
\end{figure}


\begin{thebibliography}{999}
\bibitem{dsh}''Degree of superheat'' is generally defined as the 
difference between the vapor pressure at the operating temperature 
and the operating pressure, i.e., is a measure of how far from 
equilibrium the compound is, and therefore of instability or 
sensitivity to radiation. 
CF$_{3}$I is only moderately superheated at room 
temperature, with a vapor pressure $\sim$50 psig, leading to the need 
to run 
the chamber at slightly higher temperatures (30-40 C) to achieve 
thresholds in recoil energy as low as few keV.
\bibitem{suff}In shielded underground conditions this weak neutron source is 
sufficient to increase the present bulk
bubble nucleation rate by a factor of four.
\bibitem{soft}The classical theory of radiation-induced boiling 
\cite{seitz,bell} predicts, perhaps naively, sharp (step function) 
bubble nucleation thresholds. In contrast to \cite{picasso}, where soft 
(sigmoidal) thresholds are invoked to interpret neutron 
calibrations, we have observed no 
need to deviate from the predictions of this theory. 
In dedicated experiments using 
smaller devices and neutron-induced recoils we 
have ascertained that, at least for a CF$_{3}$I bubble chamber, 
thresholds are very close to step-like. The rapid onset of full 
response to alpha-recoils observed in this larger chamber ({\frenchspacing 
Fig. 5}) is another indication of the same.
Adoption of a sigmoidal threshold would result in 
an enhanced WIMP sensitivity, most noticeably for low WIMP masses.
\end{thebibliography}
\end{document}